\documentclass[aps,prd,reprint,groupedaddress,nofootinbib]{revtex4-2}

\usepackage[T1]{fontenc}
\usepackage[utf8]{inputenc}
\usepackage{microtype}
\usepackage{amsmath,amssymb,mathtools}
\usepackage{siunitx}
\usepackage{graphicx}
\usepackage{hyperref}
\hypersetup{colorlinks=true, urlcolor=blue, citecolor=blue, linkcolor=blue}

\newcommand{\e}{\varepsilon}               % kinetic mixing
\newcommand{\rhodm}{\rho_{\mathrm{DM}}}
\newcommand{\mX}{m_{X}}                    % dark-photon mass
\newcommand{\fX}{f_{X}}                    % dark-photon frequency
\newcommand{\QL}{Q_{\!L}}                  % loaded Q
\newcommand{\Tint}{T_{\mathrm{int}}}       % integration time
\newcommand{\Tsys}{T_{\mathrm{sys}}}       % system noise temp

\newcommand{\betac}{\beta}                 % coupling coefficient
\newcommand{\CL}{\%\,\mathrm{C.L.}}        % confidence level

\begin{document}

\title{Follow-up Search for a Tentative Dark Photon Signal Near \SI{19.5}{\mu eV} using ORGAN-Q infrastructure}

\author{Aaron P.~Quiskamp}
\author{Graeme R.~Flower}
\author{Maxim Goryachev}
\author{Michael E.~Tobar}
\affiliation{Quantum Technologies and Dark Matter Research Laboratory,
  University of Western Australia, 35 Stirling Highway, Crawley WA 6009, Australia}

\author{Ben T.~McAllister}
\email{bmcallister@swin.edu.au} % first-page footer note automatically
\affiliation{Centre for Astrophysics and Supercomputing,
  Swinburne University of Technology, John St, Hawthorn VIC 3122, Australia}

\collaboration{The ORGAN Collaboration}\noaffiliation

\date{\today}

\begin{abstract}
A recent independent dark photon (DP) focused reanalysis of existing data from the TASEH axion haloscope experiment reported a tentative DP dark matter signal with local significance $\sim 4.7\sigma$ at $\fX \simeq \SI{4.71}{GHz}$, corresponding to $\mX \simeq 19.5~\mu$eV and kinetic mixing $\e \sim 6.5\times 10^{-15}$. Motivated by this report, we performed a dedicated, narrowband follow-up experiment to confirm or refute the signal with a cryogenic microwave cavity operated \emph{without} a magnetic field, leveraging the ORGAN-Q dilution refrigeration and receiver chain. Scanning a window centered on the reported frequency over a live time of $\Tint\sim 13$ days, we find no excess consistent with a dark photon signal as reported, and set $95\CL$ exclusion on $\e$ in a narrow mass range around $\sim 19.5~\mu$eV, excluding a signal of the strength and frequency reported to 99.92$\%$ confidence. We discuss the experiment and present the exclusion limits. 
\end{abstract}

\maketitle

\section{Introduction}\label{sec:intro}
The nature of dark matter remains one of the biggest mysteries in modern science~\cite{DM1,DM2,DM3}. One candidate for dark matter is an ultralight vector particle known as a ``dark photon'' (DP), which kinetically mixes with the Standard Model photon, producing an oscillatory effective electric field. The key parameters of the DP such as its mass ($\mX$) and the strength of its coupling to the Standard Model photon ($\e$), are not known, creating a wide parameter space which experiments must search~\cite{DPBackground1,DPBackground2,DPBackground3,DPBackground4,DPBackground5,DPBackground6}.
Cavity haloscopes are a kind of experiment (typically) consisting of a tunable, cryogenic, high-\(Q\) electromagnetic resonant cavity coupled to a sensitive electromagnetic field readout designed to probe the photon power inside the cavity~\cite{Haloscope1,Haloscope2,Haloscope3}. Sometimes these experiments are constructed specifically to search for dark photons~\cite{DPHalo1,DPHalo2,DPHalo3}, and sometimes they are primarily designed to search for a different kind of dark matter, the axion. In either case typical haloscopes are natural DP detectors since the DP-induced effective electric field can drive cavity resonant modes in a similar way to the axion, and thus DPs can be detected or excluded with appropriate reanalysis of existing axion haloscope data~\cite{HaloscopeDPLimits1,HaloscopeDPLimits2}.
Some axion haloscope experiments have published such analyses explicitly~\cite{ORGANDP}, and others have been recast by independent analyses~\cite{AxionLimits}.
Recently, a DP focused reanalysis of existing axion haloscope data from the TASEH experiment, conducted by an independent group, reported a tentative signal consistent with DP dark matter at $\fX\approx4.71$GHz, with $\e \sim 6.5\times 10^{-15}$~\cite{TASEH2025}. 
To verify or refute this signal, we conducted an independent follow-up experiment, using a purpose-built narrow-tuning cavity haloscope at milliKelvin temperatures, employing the cryostat and low-noise receiver chain of ORGAN-Q, an axion haloscope experiment targeting a similar frequency range to TASEH, but no magnetic field. 
This paper summarizes the apparatus, search, analysis, and limits from that targeted follow-up campaign.

\section{Dark Photon Haloscopes}\label{sec:theory}
A DP field of mass $\mX$, which kinetically mixes with the Standard Model photon with some mixing parameter $\e$ provides a source of photons with a frequency $\fX$ proportional to the DP mass, such that
\begin{equation}
    h\fX = \mX c^2 + \frac{1}{2}\mX v_{X}^2, 
    \label{eq:fsig}
\end{equation}

\begin{figure}[t]
    \centering
    \includegraphics[width=0.8\linewidth]{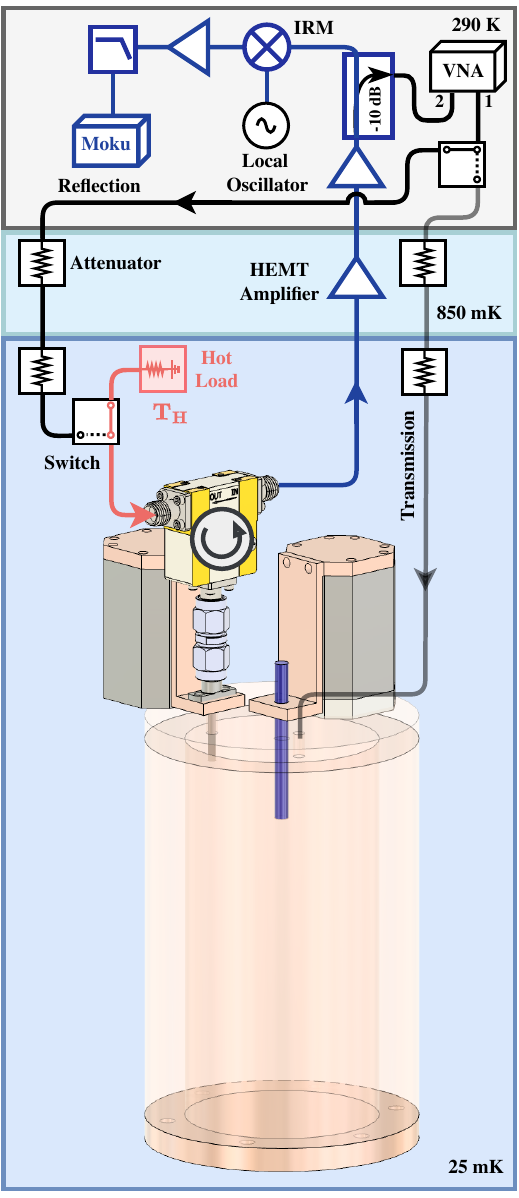}
    \caption{Schematic of the experimental setup includes a cryogenic resonant cavity at 25 mK, coupled to a low-noise HEMT amplifier chain for signal readout (blue line). The amplified signal is mixed down using an image-rejection mixer (IRM) before being digitized by a Moku Pro. A variable-temperature hot load and switch enable calibrated noise-injection for system noise characterisation. Microwave transmission and reflection paths are measured using a vector network analyzer (VNA). Cryogenic piezoelectric actuators (grey) control movement of the antenna (coupled directly to a cryogenic circulator) and the sapphire tuning stub (blue).}  
    \label{fig:schematic}
\end{figure}

where $h$ is Planck's constant, and $v_X$ is the velocity of the DP. If the DP is dark matter, the velocity distribution of dark matter particles creates a photon signal with a small effective line-width in frequency space. For typical cold dark matter models, it is assumed that $\Delta\fX\approx 10^{-6}\fX$, which can equivalently be defined as a dark photon effective `quality factor', $Q_{\text{DP}}\approx10^6$~\cite{DMSpread}.

A dark photon haloscope works on the principle that this DP induced field can drive electromagnetic resonant cavity modes, and the resulting power deposited in the cavity from DP conversion can be read out by coupling to a receiver chain. The signal power depends on a range of both cavity and DP parameters.

When the frequency of a resonant mode of the cavity is tuned near the DP induced electromagnetic field frequency, the expected DP conversion signal power that can be extracted from the cavity into a receiver chain is given by~\cite{HaloscopeDPLimits2}
\begin{equation}
    P_{dp} = \epsilon^2 m_X \rho_{\text{DM}} V G Q\frac{\beta}{1+\beta},
    \label{eq:Psig}
\end{equation}
where $\rho_{\text{DM}}$ is the local density of dark matter, $V$ is the volume of the cavity, $Q = \min(Q_L,Q_{\text{DP}})$ (with $Q_L$ being the loaded cavity quality factor), $\beta$ is the coupling coefficient of the cavity to the receiver chain, and $G$ is a dimensionless form factor which quantifies the overlap between the electric field of the cavity mode, $\vec{E}$ and the dark photon field, $\vec{X}$. It is given by
\begin{equation}
G=\frac{\left|\int \vec{E}\cdot\vec{X} dV\right|^2}{V\int\left|\vec{E}\right|^2 dV}.
\end{equation}
For the resonant modes typically employed in haloscope searches (and for the search described here), the TM$_{0n0}$ modes, $G$ can be reduced to
\begin{equation}
G=\frac{\left|\int \vec{E}_z dV\right|^2}{V\int\left|\vec{E}\right|^2 dV}\times\left<\cos^2(\theta)\right>_T,
\end{equation}
where $\vec{E}_z$ is the z-component of the cavity mode electric field, $\theta$ is the polarisation angle between the dark photon field and the cavity $z$-direction, and $\left<\cos^2(\theta)\right>_T$ is the time averaged value of $\cos^2\theta$ over the time of the search.

The relevant value of $\cos^2\theta$ for a given search depends on the model for the polarisation of the dark photon field being considered, as well as the location and duration of the search. For the random polarisation scenario assumed in the TASEH reanalysis, $\left<\cos^2(\theta)\right>_T = 1/3$~\cite{HaloscopeDPLimits1}.

In a DP haloscope, the cavity is tuned to some frequency, coupled to a low-noise readout chain, and data is acquired for some amount of time. Then, the cavity is tuned to a new frequency and more data is taken. Eventually, the presence or lack of a DP signal above the noise background of the receiver chain is used to either claim detection, or exclude DP dark matter in a mass range corresponding to the frequency range of the search, at a coupling level corresponding to the sensitivity of the search.\\

\section{ORGAN-Q TASEH Follow Up Experiment}\label{sec:apparatus}
\begin{figure*}[t]
    \centering
    \includegraphics[width=\textwidth]{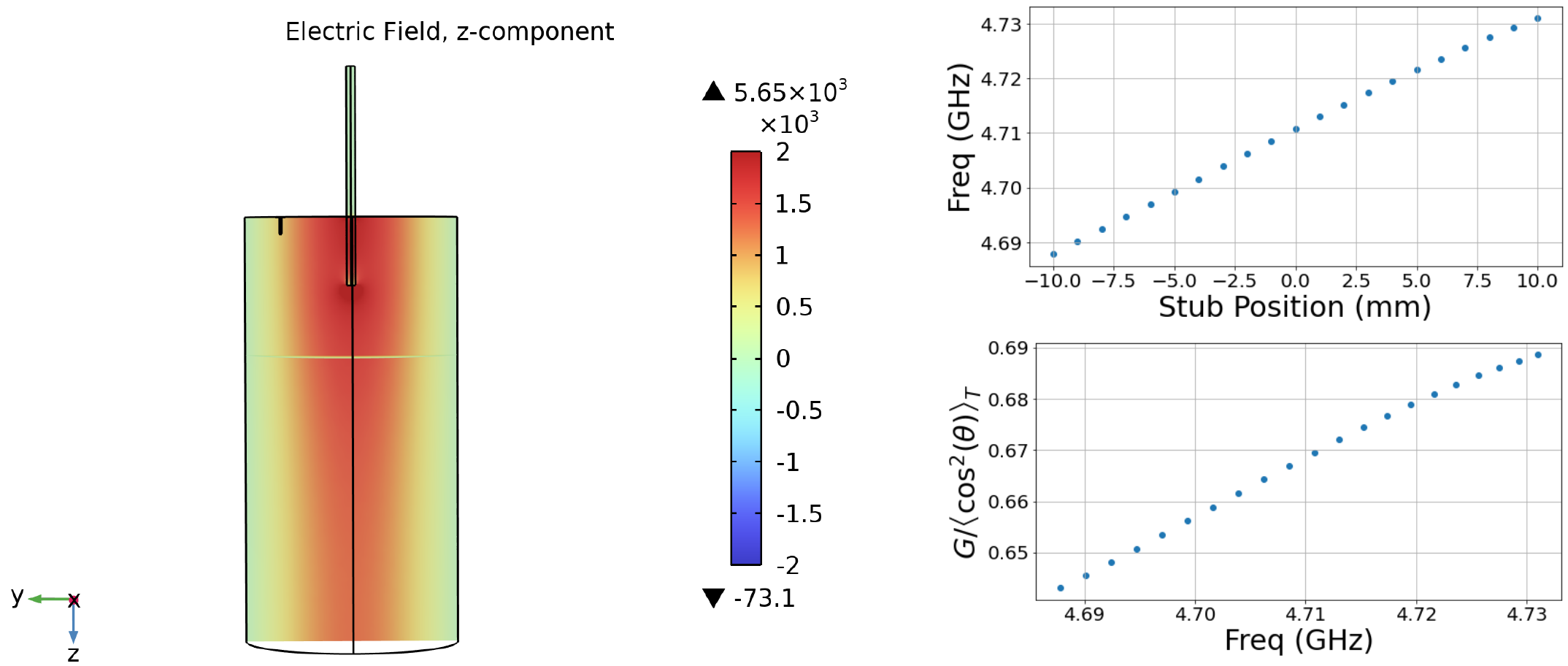}
    \caption{Left: simplified COMSOL simulation of the cavity used in the experiment. The tuning stub, and strongly coupled antenna can be seen. The simulation was used to inform the sensitivity, and compared against experimental results. Top right: the simulated frequency against tuning stub position along the z-axis. The `0' position was set such that the mode was at the frequency of interest, and the cavity was tuned around this position. Bottom right: the simulated geometry factor divided by the polarisation angle normalisation against frequency. The polarisation normalised geometry factor presented here is equivalent to the form factor typically presented in axion haloscopes.}
    \label{fig:design}
\end{figure*}

TASEH is an axion haloscope experiment, which in 2022 excluded a region of axion dark matter space around $19.5~\mu$eV~\cite{TASEHAxion2022}. In excluding axion dark matter, the TASEH search employed axion-specific vetoes to discount candidate signals. 

As mentioned, a recent independent reanalysis of the TASEH data in the dark photon context, not using these axion specific vetoes, reported a tentative signal with $\mX\approx19.5\mu eV$, and $\e\approx 6.5\times 10^{-15}$~\cite{TASEH2025}. The ORGAN Collaboration is an axion haloscope experimental collaboration which has placed limits on axions and dark photons across a variety of mass ranges from a range of haloscope experiments~\cite{ORGAN1,ORGAN2,ORGAN3}. ORGAN-Q is one of the experiments within The ORGAN Collaboration, and it searches for axions around 6 GHz~\cite{ORGANQ}. The infrastructure that supports ORGAN-Q, such as the dilution refrigerator, readout chain and DAQ can all in principle be used for other searches in similar frequency ranges, such as the $\sim$4.7 GHz range of the TASEH search. In light of this, we constructed a dedicated follow-up resonant cavity experiment using the ORGAN-Q infrastructure to confirm or refute this tentative DP signal.

The follow-up detector comprised a high-Q cylindrical copper cavity, designed to tune over a narrow frequency range with the co-axial insertion a sapphire tuning stub. The cavity was specifically designed to target the frequency range of the tentative TASEH reanalysis signal. The tuning stub was positioned and adjusted inside the cavity with a cryogenic piezoelectric actuator. The cavity was designed using simulations in COMSOL Multiphysics, taking into account the cryogenic contraction of the materials, to target the frequency range in question. Comparison between the simulations and the experimental data was used to determine how much of the sapphire stub was inserted into the cavity at a given frequency, and thus extract the expected cavity geometry factor. Given the relatively simple nature of the cavity compared to typical axion haloscopes, particularly the relatively small deviation in geometry from a perfect, empty cylinder -- the experimental and simulated parameters were in excellent agreement.

The cavity was mounted on the mixing chamber plate of the ORGAN-Q dilution refrigerator and cooled to milli-Kelvin temperatures. No magnetic field was applied. Two antennae were inserted into the cavity, one was weakly coupled for transmission measurements. The second, strongly coupled antenna was connected via a cryogenic circulator to a first-stage HEMT amplifier mounted at the 1 K plate of the refrigerator. The signal from the output of the HEMT passed through additional amplification, and was mixed down to lower frequencies for acquisition with an FPGA-based digitizer. A Y-factor measurement setup was employed to calibrate the noise temperature of the readout chain.

A schematic of the experiment and readout chain can be seen in Fig.~\ref{fig:schematic}, and outputs from the cavity design simulations can be seen in Fig.~\ref{fig:design}. 

\begin{figure*}[t]
    \centering
    \includegraphics[width=\textwidth]{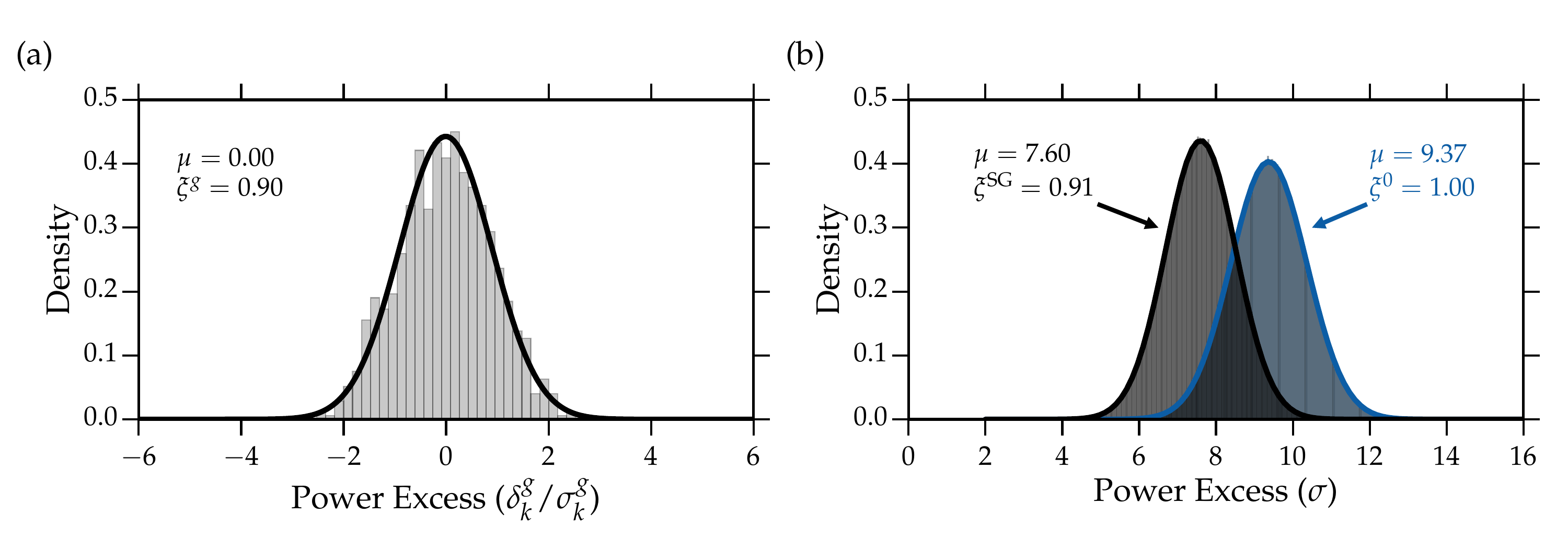}
    \caption{(a) Distribution of normalised grand spectrum bins $\delta^g_k/\sigma^g_k$ following Savitzky–Golay (SG) filtering, vertical combination of overlapping spectra, and horizontal re-binning according to the Maxwell–Boltzmann signal template. The observed narrowing, quantified by $\xi^g = 0.90$, is attributed to the negative correlations introduced by the SG filter. (b) Signal-to-noise ratio (SNR) histograms for a synthetic DP signal embedded in Gaussian white noise, with one dataset passed through the SG filter (black) to simulate the analysis pipeline and the other left unfiltered (blue). The corresponding attenuation of a DP signal is quantified by the ratio of the mean SNRs, $\eta^{\text{SG}} = 7.60/9.37 = 0.81$.}
    \label{fig:histogram}
\end{figure*}

The typical key parameters of the search are summarised here:
\begin{itemize}
    \item Target mode: $\mathrm{TM}_{\mathrm{010}}$.
    \item Frequency: $\nu_0 \approx 4.71018$ GHz.
    \item Cavity dimensions: radius 24.3 mm, height 97.2 mm.
    \item Sapphire stub dimensions: radius 1.76 mm.
    \item Loaded quality factor: $\QL =$ 19700.
    \item Coupling: $\betac=$1.06.
    \item Cavity volume: 180.3 mL.
    \item System noise temperature: $\Tsys=$ 3.33 K
    \item Total integration time: $\sim 13$ days.
    \item Resolution bandwidth: 397 Hz.
\end{itemize}

\section{Data Taking and Calibration}\label{sec:run}
We collected $\sim$13 live days of data in August 2025. The sapphire stub tuner stepped the resonance so that the resonant mode bandwidth covered a range around the reported DP signal frequency of 4.71018 GHz. The antenna coupling was adjusted to $\beta\simeq1$ to maximise the DP signal power whilst minimising resonance distortion or mixing of the resonant mode with the standing waves in the readout circuit. 

We measured a narrowband, high-resolution power spectrum and performed \emph{in situ} VNA measurements of the resonance to extract $\QL$, $\betac$, and frequency. The total system noise temperature $\Tsys$ was measured using a Y-factor method via a cryogenic switch, as shown in Fig. \ref{fig:schematic}. In this configuration, the cryogenic termination (load) was mounted on a copper block equipped with a calibrated temperature sensor and a resistive heater. The block was mechanically anchored to the mixing chamber plate through a 50 mm stack of alumina washers, which provided a weak thermal link. The load was heated from 170 mK to 3.32 K in eight evenly spaced steps, and the corresponding output noise powers were recorded. A linear fit of output power versus effective input temperature $T_\text{H}$, gave the total system noise. To refer this noise to the cavity output, we measured the attenuation between the heated load setup (red line in Fig. \ref{fig:schematic}) and the circulator output (the cavity input) during a separate, dedicated cooldown. From this calibration, the attenuation $\alpha_Y$, was found to be $\sim 1.03 \pm 0.07$ dB, and the mean value of $\Tsys$ in the search region was calculated to be 3.33 K. 

\section{Analysis}\label{sec:analysis}

\begin{figure*}[t]
    \centering
    \includegraphics[width=0.8\textwidth]{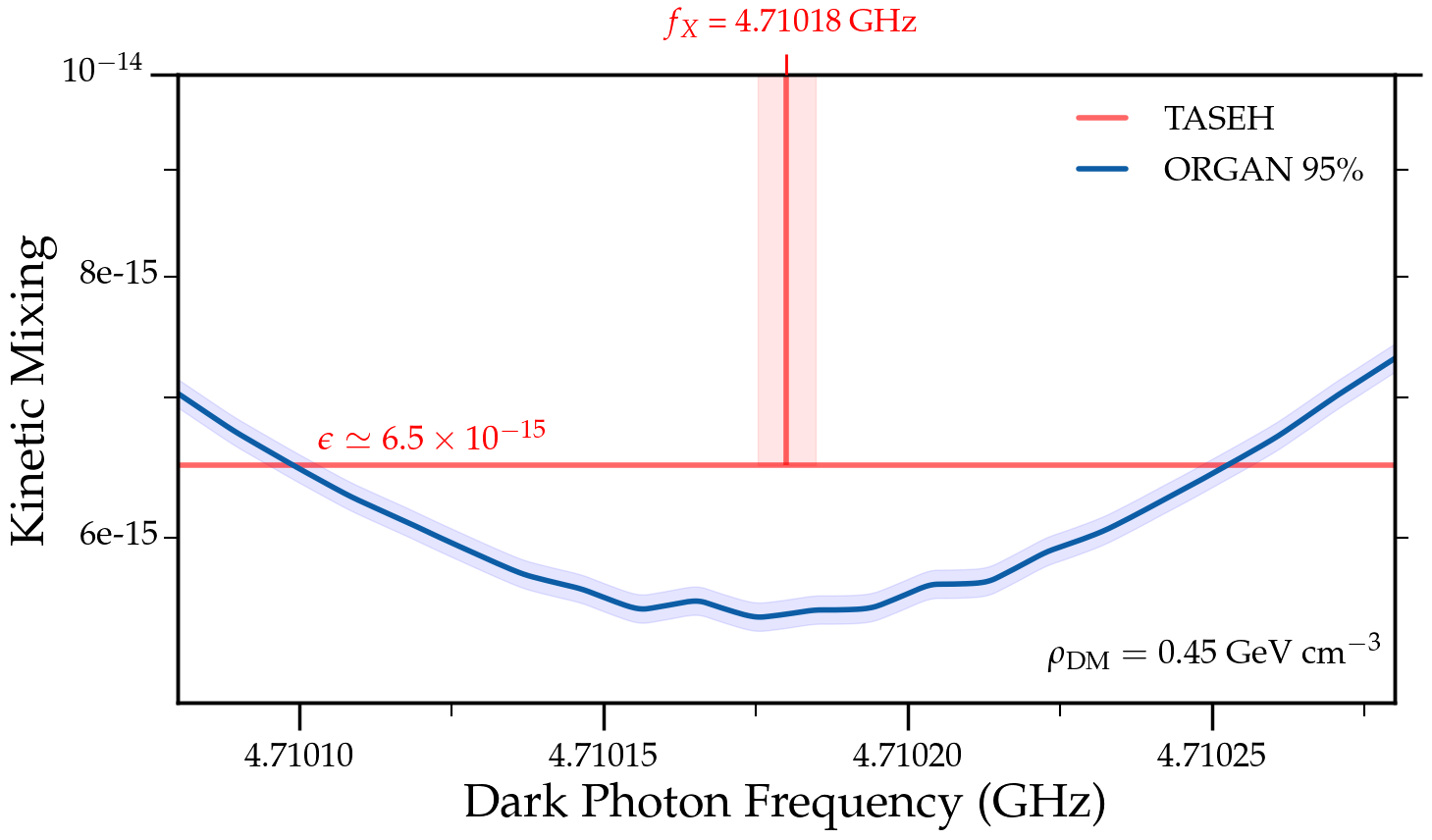}
    \caption{Exclusion limits on the dark photon kinetic mixing parameter \(\epsilon\) for the ORGAN follow-up experiment assuming the random polarization scenario. The 95\% confidence level limits are shown in blue, assuming a local dark matter density of \(\rho_{\text{DM}}=0.45\,\text{GeV cm}^{-3}\). The shaded band represents the associated uncertainty in the limit. For comparison, the claimed signal reported by the TASEH collaboration is shown in red, with the horizontal line denoting the reported mixing strength of \(\epsilon\simeq6.5\times10^{-15}\) and the vertical band marking the signal frequency \(\pm 1\) expected dark photon linewidth.}
    \label{fig:limits}
\end{figure*}

With the collected data, we perform a standard haloscope dark matter signal search, adapted from the HAYSTAC procedure~\cite{HAYSTACAnalysis}. Since all data were collected at a fixed cavity tuning, the usual vertical combination reduces to a simple average of the traces. The resulting spectrum is normalised by dividing out a smoothed baseline obtained from a Savitzky–Golay filter, removing the frequency-dependent gain and cavity response, while leaving narrow-band features intact. The flattened spectra are scaled according to the measured system noise temperature and the expected dark-photon signal power derived from the experimental parameters. 

In this analysis we applied a maximum-likelihood filter matched to the expected dark matter signal lineshape.  we adopt the Standard Halo Model, assuming the dark matter velocity distribution follows a boosted Maxwell–Boltzmann profile with an effective dispersion that accounts for Earth’s motion through the halo~\cite{HAYSTACAnalysis,DMSpread}. Horizontal re-binning was done by summing groups of 12 consecutive bins, chosen to match the expected dark matter signal width. The resulting normalized grand power spectrum is expressed as \(\delta^g_k / \sigma^g_k\), where \(\delta^g_k\) denotes the power excess in the \(k\)-th grand spectrum bin and \(\sigma^g_k\) is the corresponding standard deviation of the noise. In the absence of a signal, a standard normal distribution centered at zero is expected. However, due to negative correlations induced the SG-filtering step, this width is reduced to $\xi^g=0.90$. This effect was verified through a dedicated simulation in which synthetic dark photon signals were injected into Gaussian noise and processed using the same analysis pipeline. The results of \(10^6\) iterations of this simulation, shown as a histogram in Fig.~\ref{fig:histogram}, demonstrate that the signal-to-noise ratio is suppressed by a factor \(\eta^{\mathrm{SG}}=0.81\). After accounting for the additional narrowing of the grand spectrum distribution by $\xi^g$, the overall attenuation of the signal-to-noise ratio is given by \(\eta^{\text{SNR}}=0.9\).   

In this search, the candidate threshold was chosen such that no bins passed the threshold, and thus no rescans were performed. For full details of the statistical treatment, matched filtering, and rescan protocol, we refer to Refs.~\cite{ORGAN2,ORGAN3,ORGANQ,HAYSTACAnalysis}.

\section{Results}\label{sec:results}
No persistent excess compatible with the DP template was observed around $\fX$=~4.71018 GHz. 
Under the random polarization assumption adopted in the TASEH reanalysis, adopting the experimental parameters of our search, with $\rhodm$=~0.45 GeV/cm$^3$, we set 95$\CL$ upper limits of
\[
  \e \,<\, 6.5\times 10^{-15}\,,
\] 
within $\pm\sim$15 expected DP signal linewidths of the reported mass from the TASEH reanalysis.
The limits are more stringent than this over most of the range, and are presented along with the TASEH reanalysis signal properties in Fig.~\ref{fig:limits}. The uncertainties in these limits are summarized in Table~\ref{table:uncertainties}.
\begin{table}
\begin{center}
\begin{tabular}{| c | c |}
\hline
$T_{sys}$ & 1.6\% \\ 
\hline
$\beta$ & 2.4\% \\
\hline
$Q_L$ & 2.2\% \\
\hline
$G$ & 1.0\% \\
\hline
Total with respect to $\e$ & 1.6\% \\
\hline
\end{tabular}
\caption{Percentage uncertainties contributing to the uncertainty on the exclusion limits.}
\label{table:uncertainties}
\end{center}
\end{table}\\

The uncertainty in system noise temperature is derived from the uncertainties in the measured parameters used in the Y-factor model ($T_\text{H}$ and $\alpha_Y$), which are carried through the analysis. The uncertainty in $\beta$ and $\QL$ are taken as the standard errors obtained from the fits to the VNA measurements. The uncertainty in $G$ is derived by COMSOL modeling of geometric uncertainties -- both of the physical dimensions of the cavity and other components, and on the locations of the sapphire tuning stub and antennae. These uncertainties were found to be $<1\%$, but were rounded in the interest of conservative analysis. We attribute the relatively low uncertainty in $G$ to the relative simplicity of the cavity and constituent components compared to typical haloscope searches, as well as the lack of an external magnetic field -- the convolution of which with the cavity mode field adds additional uncertainty in the axion haloscope case.

\section{Discussion}\label{sec:discussion}
Our null result disfavors a universal, instrument-independent DP dark matter signal with the parameters reported by the independent TASEH reanalysis. Whilst we set broader 95$\CL$ exclusions on the DP kinetic mixing in a region of parameter space around the signal, as reported above, when considering the \emph{specific} frequency and coupling reported in the TASEH reanalysis, our result excludes a signal of that strength with 99.92$\%$ confidence.
Additionally -- this analysis is conducted in the random polarisation scenario, as that is what was considered in the TASEH reanalysis. Regardless, we can also consider the standard fixed polarisation scenario. In the fixed polarisation scenario, for the location and timing of this follow-up search, following the procedure set out in~\cite{HaloscopeDPLimits1}, we compute $\left<\cos^2(\theta)\right>_T\approx0.3157$ for the 95$\CL$ case -- meaning that in the fixed polarisation scenario, our 95$\CL$ on $\e$ is degraded by approximately 2.75$\%$. Importantly, in this scenario, the TASEH reanalysis signal strength would also be degraded -- and by a larger margin owing to the lower value of $\left<\cos^2(\theta)\right>_T$ for the location and timing of the TASEH experiment. As such, our experiment excludes the TASEH reanalysis signal in both the random and fixed polarisation scenarios. We have only presented the limits in the random polarisation scenario in Fig.~\ref{fig:limits} for readability and direct comparison to the TASEH reanalysis results, but the limits can be easily rescaled in the fixed scenario. 
\section{Conclusion}\label{sec:conclusion}
We have carried out a focused, cryogenic cavity haloscope follow-up of the tentative dark photon dark matter signal reported in an independent reanalysis of TASEH data. The tentative signal is near 4.71 GHz in frequency space. Our follow-up uses the ORGAN-Q infrastructure without a magnetic field, and a dedicated detector cavity. 
Within our scanned window and under standard dark matter halo and TASEH reanalysis polarisation assumptions we find no evidence for a dark photon and set 95$\CL$ exclusions that cover $\pm$~15 expected signal linewidths around the reported frequency, excluding the specific reported signal characteristics to 99.92$\%$ confidence. 
\begin{acknowledgments}
This work is supported by Australian Research Council Grants CE200100008 and DE250100933, as well as the Australian Government’s Research Training Program.
\end{acknowledgments}

\end{document}